\documentclass[a4paper,11pt]{article}
\pdfoutput=1 
\usepackage{jheppub}
\usepackage[T1]{fontenc} 
\usepackage{color}
\usepackage{fancybox}
\usepackage{amsmath}
\usepackage{amsfonts}
\usepackage{slashed}
\usepackage{amssymb}

\newcommand{\be}{\begin{equation}}
\newcommand{\bea}{\begin{eqnarray}}

\newcommand{\ee}{\end{equation}}
\newcommand{\eea}{\end{eqnarray}}

\title{\boldmath DYONIC $AdS_{4}$ BLACK HOLE ENTROPY AND ATTRACTORS VIA ENTROPY FUNCTION}

\author{PRIESLEI GOULART}

\affiliation{Instituto de F\'{i}sica Te\'{o}rica, UNESP-Universidade Estadual Paulista\\ R. Dr. Bento T. Ferraz 271, Bl. II, Sao Paulo 01140-070, SP, Brazil}
\affiliation{Max-Planck-Institut f\"{u}r Physik (Werner Heisenberg Institut)\\
F\"{o}hringer Ring 6, D-80805 Munich, Germany}

\emailAdd{prieslei@ift.unesp.br}
\emailAdd{prieslei@mpp.mpg.de}

\abstract{Using the Sen's entropy function formalism, we compute the entropy for the extremal dyonic black hole solutions of theories in the presence of dilaton field coupled to the field strength and a dilaton potential. We solve  the attractor equations analytically and determine the near horizon metric, the value of the scalar fields and the electric field on the horizon, and consequently the entropy of these black holes. The attractor mechanism plays a very important role for these systems, and after studying the simplest systems involving dilaton fields, we propose a general solution for the value of the scalar field on the horizon, which allows us to solve the attractor equations for gauged supergravity theories in $AdS_{4}$ spaces. In particular, we derive an expression for the dyonic black hole entropy for the $\mathcal{N}=8$ gauged supergravity in 4 dimensions which does not contain explicitly the gauge parameter of the potential. }


\begin{document} 
\hfill{MPP-2015-300}
\maketitle

\section{Introduction}
The structure of charged black holes solutions in string theory differs a lot from the usual Reissner-Nordstrom solution. Such a difference is due to a nontrivial coupling between the dilaton field and the Maxwell field $F^{\mu\nu}$ required to describe a low-energy effective Lagrangian in string theory. The Reissner-Nordstrom solution describes black holes containing both electric and magnetic charges at the same time (dyonic black hole), or just one of them separately. For low-energy Lagrangians coming from string theory the nontrivial coupling between the dilaton and the Maxwell field makes difficult to obtain analytical solutions for the resulting Einstein's equations when the black hole is dyonic. The situation gets even worse when the chosen compactification scheme introduces a dilaton potential with a complicated form, as is the case of gauged supergravities in $AdS$ spacetimes. 

The first achievements in obtaining such solutions in a closed form in the presence of the dilaton field was due to Gibbons and Maeda in \cite{Gibbons:1987ps}, and later, the same solution was independently found by Garfinkle, Horowitz and Strominger in \cite{Garfinkle:1990qj}. In both cases, the solution corresponds to a magnetically charged black hole that is related to the electrically charged black hole via a duality transformation. Later, the solution in the presence of both electric and magnetic charges was found in \cite{Shapere:1991ta}, and they were only possible to be obtained by requiring the existence of another scalar field, the axion field, that also couples non-trivially to the Maxwell field. In the presence of both the dilaton and the axion, the equations of motion are invariant under $SL(2,R)$ transformations, which are a generalization of the previous duality transformation. In the absence of an axion, the dyonic black hole solution for the $\mathcal{N}=4$ ungauged supergravity was found in \cite{Kallosh:1992ii}. In  \cite{Gibbons:1987ps, Garfinkle:1990qj, Shapere:1991ta} the black hole has the metric and thermodynamic properties of a Schwarzschild solution, whereas the solution found in \cite{Kallosh:1992ii} has the metric and thermodynamic properties of a Reissner-Nordstrom black hole.

The attractor mechanism for black holes states that the value of the scalar fields on the horizon is independent of their values at infinity, and it depends only on the charges of the black hole. It was first discovered in the context of supergravity \cite{Ferrara:1995ih, Strominger:1996kf, Ferrara:1996dd}, but later it was shown that supersymmetry does not play an important role in the attractor phenomenon, and even after the inclusion of higher derivatives terms in the action such mechanism still occurs (see for instance \cite{Sen:2005wa} and references therein), and it is believed to be a universal feature of any gravity theory containing scalars. The attractor mechanism became a so powerful tool that it allowed Sen \cite{Sen:2005wa} to develop a whole formalism to calculate not only the value of the scalar fields on the horizon but also the entropy of extremal black holes, for theories containing also higher derivative terms. The importance of this result resides in the classical and quantum aspects of the black holes, since entropy is related to the counting of their microstates. 

It is undeniable that the gauge-gravity duality \cite{Maldacena:1997re, Witten:1998qj}, not only reinforced  the interest in black hole solutions in supergravities in $AdS$ spacetimes, but also has given new insights to the study of strongly coupled systems. Applications of holographic methods to study superconductivity has recently been a subject of intense investigation. The first model for a holographic superconductor was created by Gubser in \cite{Gubser:2008px} and reviewed in \cite{Horowitz:2010gk} for instance. This theory is just gravity with a negative cosmological constant coupled to a Maxwell system and a charged massive scalar. Efforts were made to embed holographic superconductor models in string theory \cite{Denef:2009tp, Gauntlett:2009dn, Gauntlett:2009bh, Gubser:2009qm}, but the situation gets more complicated due to the nonminimal coupling of the scalar fields and also the more complicated forms of the scalar potentials coming from supergravity. Although the scalar fields considered in this paper are not charged and not minimally coupled, we believe that understanding the near horizon properties of these complicated systems might be a crucial step towards a better description of not only the counting of microstates of the black hole but also of the holographic superconductors, as the infrared properties are important to figure out phenomenologically viable models for superconductivity. Also, from the gauge-gravity duality point of view, the moduli flow is interpreted as an RG flow \cite{Goldstein:2005hq, Astefanesei:2007vh}.

In the context of $\mathcal{N}=2$ gauged supergravity, black holes in $AdS_{4}$ spaces have been extensively studied, and supersymmetric black hole solutions with running scalar fields were found in \cite{Cacciatori:2009iz}, and explored with more details in \cite{Dall'Agata:2010gj, Halmagyi:2013qoa, Halmagyi:2014qza}. Extensions for the non-supersymmetric case can be found for instance in \cite{Toldo:2012ec, Gnecchi:2012kb, Hristov:2013sya, Gnecchi:2014cqa}, and references therein. 

In this work we compute the entropy for extremal black holes with spherical symmetry using the entropy function formalism, by assuming that the near horizon geometry of these theories is $AdS_{2}\times S^{D-2}$.  The paper is organized as follows. In section 2 we derive the equations of motion for a general dilaton theory in which the dilaton couples nontrivially to the Maxwell field and has a dilaton potential. In section 3 we review the Sen's entropy function formalism and write the attractor equations. In section 4 we apply this formalism to compute the entropy for the extremal black hole for different potentials. We derive the near horizon metric, the value of the scalars on the horizon, the electric field and the entropy follows directly. In section 5 we apply this for $AdS_{4}$ black holes, and we speculate about the generality of the ansatz for the dilaton on the horizon, which allowed us to determine explicitly the solution for the attractor equations and the entropy of this dyonic black hole. In section 6 we present a summary and conclusions.

\section{The dilaton action}
The general dilaton action in the presence of a scalar potential is
\be S=\int d^{4}x\sqrt{-g}\left(R-2\partial_{\mu}\phi\partial^{\mu}\phi-W(\phi)F_{\mu\nu}F^{\mu\nu}-V(\phi)\right). \label{ad}\ee
Here, we define the field strength as
\be F_{\mu\nu}=\partial_{\mu}A_{\nu}-\partial_{\nu}A_{\mu}, \ee
and we take units in which $16\pi G=1$. The action is written in terms of a function of the dilaton field, $W(\phi)$, coupled to the field strength,  and a dilaton potential $V(\phi)$. This is a general form of the bosonic part of some gauged supergravity theories, where  $W(\phi)$ is in general an exponential function of the dilaton, and $V(\phi)$ is a generic potential whose form may vary due to the choice of the gauge for scalar fields in truncated supergravity theories. One could also have more than one scalar or gauge field, as the case of $U(1)^{4}$ gauged supergravity described in the text and in the appendix, but all the formulas obtained are easily generalized for these cases.

The equations of motion are:
\begin{itemize}
\item for the metric:
\be R_{\mu\nu}=2\partial_{\mu}\phi\partial_{\nu}\phi-\frac{1}{2}g_{\mu\nu}W(\phi)F_{\rho\sigma}F^{\rho\sigma}+2W(\phi)F_{\mu\rho}
{F_{\nu}}^{\rho}+\frac{1}{2}g_{\mu\nu}V(\phi); \ee
\item for the dilaton:
\be \nabla_{\mu}(\partial^{\mu}\phi)-\frac{1}{4}\frac{\partial W(\phi)}{\partial\phi}F_{\mu\nu}F^{\mu\nu}-\frac{1}{4}\frac{\partial V}{\partial \phi}=0; \ee
\item for the gauge field:
\be \nabla_{\mu}\left(W(\phi)F^{\mu\nu}\right)=0. \ee
We also have the Bianchi identities:
\be \nabla_{\left[\mu\right.}F_{\left.\rho\sigma\right]}=0.\ee
\end{itemize}
\section{Entropy function and attractor equations}

The basic assumption we have to make to compute the Sen's entropy function is that the spherically symmetric extremal black hole solution has the near 
horizon metric given by
\be 
ds^{2}=v_{1}\left(-r^{2}dt^{2}+\frac{dr^{2}}{r^{2}}\right)+v_{2}(d\theta^{2}+\sin^{2} \theta d\phi^{2}), \label{one} 
\ee
where the constants $v_{1}$ and $v_{2}$ are the $AdS_{2}$ radius and the $S^{2}$ radius respectively. This means that we are assuming that the near horizon geometry is $AdS_{2}\times S^{2}$. It is believed that an $AdS_{2}\times S^{D-2}$ is a general feature of extremal black holes with spherical symmetry in $D$ dimensions, and this was proven in $4$ and $5$ spacetime dimensions \cite{Kunduri:2007vf}.  The scalar and vector 
fields are constants for this geometry and are written as \cite{Sen:2005wa}
\be 
\phi_{s}=u_{s}, \,\,\, F^{(A)}_{rt}=e_{A}, \,\,\, F_{\theta\phi}^{(A)}=\frac{p_{A}}{4\pi}\sin \theta, 
\ee
where ${e_{A}}$ and ${p_{A}}$ are related to the integrals of the magnetic and electric fluxes, which are in turn related to the electric and 
magnetic charges respectively. The metric (\ref{one}) has the $SO(2,1)\times SO(3)$ symmetry of $AdS_{2}\times S^{2}$. The 
function $f(u_{s}, v_{i}, e_{A}, p_{A})$  is defined as the Lagrangian density $\sqrt{- \det g}{\mathcal{L}}$ evaluated for the near horizon geometry 
(\ref{one}) and integrated over the angular variables, 
\be 
f(u_{s}, v_{i}, e_{A}, p_{A})=\int d\theta d\phi \sqrt{- \det g}{\mathcal{L}}. 
\ee
We extremize this function with respect to $u_{s}$, $v_{i}$ and $e_{A}$ by
\be 
\frac{\partial f}{\partial u_{s}}=0, \,\,\, \frac{\partial f}{\partial v_{i}}=0, \,\,\,  \frac{\partial f}{\partial e_{A}}=q^{A},
\ee
where the first two equations are the equations of motion for the scalar and the metric respectively\footnote{Notice that in the absence of axionic couplings in the action, i.e. $aF\tilde{F}$, where $\tilde{F}$ is the dual of $F$, $q^{A}$ is simply interpreted as the electric charge. In the symplectic covariant formalism of $\mathcal{N}=2$,  $q^{A}$ has the same meaning as $q_{\Lambda}$ defined in equation 2.5 of reference \cite{Halmagyi:2013qoa} for instance. For more details, see Sen's review \cite{Sen:2007qy}.}. Next, one defines
the entropy function,
\be 
{\mathcal{E}}(\vec{u},\vec{v},\vec{e},\vec{q},\vec{p})\equiv 2\pi[e_{A}q^{A}-f(\vec{u},\vec{v},\vec{e},\vec{p})]. \label{entrfct}
\ee
The equations that extremize the entropy function are
\be 
\frac{\partial {\mathcal{E}}}{\partial u_{s}}=0, \,\,\, \frac{\partial {\mathcal{E}}}{\partial v_{1}}=0, \,\,\, \frac{\partial {\mathcal{E}}}{\partial v_{2}}=0, 
\,\,\, \frac{\partial {\mathcal{E}}}{\partial e_{A}}=0\;, \label{attracteqs}
\ee
and are called the attractor equations. At the extremum \eqref{attracteqs}, this new function equals the entropy of the black hole
\be 
S_{BH}={\mathcal{E}}(\vec{u},\vec{v},\vec{e},\vec{q},\vec{p}).
\ee

\section{Entropy function for the dilaton action with a potential}
In this section we compute the entropy of the extremal black holes considered in the text. The theories we consider do not make use of any explicit functional form for the coupling of the dilaton field with the field strength, i.e. $W(\phi)$. Also, with the exception of the constant potential, the potentials $V(\phi)$ we consider here are written in terms of these couplings in a specific way, which leads us to make generic assumptions about the solutions to the attractor equations for the case of $U(1)^{4}$ theories considered in the next section.

The quantities of interest for us are the Riemann tensor
\begin{eqnarray}
R_{\alpha\beta\gamma\delta}=-v_{1}^{-1}(g_{\alpha\gamma}g_{\beta\delta}-g_{\alpha\delta}g_{\beta\gamma}), \,\,\, 
\alpha, \beta, \gamma, \delta=r, t \nonumber \\
R_{mnpq}=v_{2}^{-1}(g_{mp}g_{nq}-g_{mq}g_{np}), \,\,\, m, n, p, q=\theta, \phi,
\end{eqnarray}
and, from it we can obtain easily the Ricci scalar
\be R=-\frac{2}{v_{1}}+\frac{2}{v_{2}}. \ee
We define the components of the field strength as $F_{rt}=e$ and $F_{\theta\phi}=P\sin\theta$.  Also, as mentioned before, $v_{1}$ is the $AdS_{2}$ radius, $v_{2}$ is the $S^{2}$ radius, and here we define the value of the scalar on the horizon as $u_{D}$. Following the procedure for obtaining the entropy function we plug our ansatze for the metric and fields in the Lagrangian, and then we integrate it over the angular variables to obtain 
\be f=4\pi v_{1}v_{2}\left[-\frac{2}{v_{1}}+\frac{2}{v_{2}}+W(u_{D})\left(-\frac{e^{2}}{v_{1}^{2}}+\frac{P^{2}}{v_{2}^{2}}\right)-V(u_{D})\right],  \ee
and the entropy function is just ${\mathcal{E}}=2\pi[Qe-f]$, which gives
\begin{eqnarray}
{\mathcal{E}}=2\pi\left[Qe-8\pi(v_{1}-v_{2})-4\pi v_{1}v_{2}W(u_{D})\left(\frac{e^{2}}{v_{1}^{2}}-\frac{P^{2}}{v_{2}^{2}}\right)+4\pi v_{1}v_{2}V(u_{D})\right].
\label{neef}\end{eqnarray} 
Taking the proper derivatives, we obtain the attractor equations
\be Q-8\pi v_{1}v_{2}W(u_{D})\frac{e}{v_{1}^{2}} =0,  \label{a1}\ee
\be -2+v_{2}W(u_{D})\left(\frac{e^{2}}{v_{1}^{2}}+\frac{P^{2}}{v_{2}^{2}}\right)+v_{2}V(u_{D})=0, \label{a2} \ee
\be 2-v_{1}W(u_{D})\left(\frac{e^{2}}{v_{1}^{2}}+\frac{P^{2}}{v_{2}^{2}}\right)+v_{1}V(u_{D})=0,  \label{a3}\ee
\be \frac{\partial W(u_{D})}{\partial u_{D}}\left(\frac{e^{2}}{v_{1}^{2}}-\frac{P^{2}}{v_{2}^{2}}\right)-\frac{\partial V(u_{D})}{\partial u_{D}}=0. \label{a4}\ee
One can check that these equations are the equations of motion derived in the previous section for the near horizon geometry. Using (\ref{a1}) we can eliminate the term containing $Q$ in (\ref{neef}), and the result is
\be
{\mathcal{E}}=2\pi\left[-8\pi(v_{1}-v_{2})+4\pi v_{1}v_{2}W(u_{D})\left(\frac{e^{2}}{v_{1}^{2}}+\frac{P^{2}}{v_{2}^{2}}\right)+4\pi v_{1}v_{2}V(u_{D})\right].
\label{neef1}\ee
Multiplying (\ref{a2}) by $v_{1}$ and adding to (\ref{a3}) multiplied by  $v_{2}$ we have a formula for the potential
\be v_{1}v_{2}V(u_{D})=(v_2-v_1) \label{potential}. \ee
This formula holds for general scalar fields $\phi^{I}$ and gauge fields $A_{\mu}^{I}$, and we will also use it in the next section. We can eliminate the potential from (\ref{neef1}) to obtain
\be
{\mathcal{E}}=2\pi\left[-4\pi(v_{1}-v_{2})+4\pi v_{1}v_{2}W(u_{D})\left(\frac{e^{2}}{v_{1}^{2}} +\frac{1}{v_{2}^{2}}\frac{p^{2}}{(4\pi)^{2}}\right)\right].
\label{neef2}\ee
Now, multiplying  (\ref{a2}) by $v_{1}$ and subtracting (\ref{a3}) multiplied by $v_{2}$ we have
\be v_{1}v_{2}W(u_{D})\left(\frac{e^{2}}{v_{1}^{2}} +\frac{P^{2}}{v_{2}^{2}}\right)=(v_1+v_2), \label{termdilaton}\ee
and eliminating this term from (\ref{neef2}) we have the entropy written only in terms of the $S^2$ radius,
\be
{\mathcal{E}}=16\pi^{2} v_{2}.
\label{neef3}\ee
One can recover the correct constant by dimensional analysis, meaning $16\pi G\equiv 1$, and show that this corresponds to the usual $A/4$ term of the black hole entropy.
Notice that we can solve (\ref{a1}) directly, giving
\be \frac{e}{v_{1}}=\frac{Q}{8\pi v_{2}W(u_{D})}. \ee
We will analyze the solutions of the attractor equations for some specific potentials, by making some considerations about its functional dependence on the dilaton field. We list the solutions case by case.
\begin{itemize}
\item $V(\phi)=0$.
\end{itemize}
For this case (\ref{a4}) gives directly the value of the function $W(u_{D})$ on the horizon, and the constants $v_{1}$ and $v_{2}$ and the entropy are
\be W(u_{D})=\frac{Q}{8\pi P},\,\, v_{1}=v_{2}=\frac{QP}{8\pi},\,\, e=P, \ee 
\be {\mathcal{E}}=2\pi QP.  \ee
This means that the value of the electric field is equal to the value of the magnetic charge  on the horizon. Again, one should notice that this result is independent of the functional form of $W(u_{D})$. This should be contrasted to the GHS solution \cite{Garfinkle:1990qj}, in which the dilaton coupling to the field strength is $W(\phi)=e^{-2\phi}$. The same result for the entropy was derived before by other method in \cite{Kallosh:1992ii}.

\begin{itemize}
\item $V(\phi)=2\Lambda$.
\end{itemize}
The solution for the attractor equations for this case is 
\begin{equation} W(u_{D})=\frac{Q}{8\pi P},\,\, e=P\frac{v_{1}}{v_{2}},\end{equation}
\begin{equation} v_{1}=\frac{1}{\Lambda}\frac{\frac{1}{2\Lambda}\left(1\pm\sqrt{1+\frac{\Lambda QP}{2\pi}}\right)-\frac{QP}{8\pi}}{\frac{QP}{4\pi}-\frac{1}{2\Lambda}\left(1\pm\sqrt{1+\frac{\Lambda QP}{2\pi}}\right)},\,\,
v_{2}=\frac{1}{2\Lambda}\left(1\pm\sqrt{1-\frac{\Lambda QP}{2\pi}}\right), \end{equation}
\be {\mathcal{E}}=\frac{8\pi^{2}}{\Lambda}\left(1-\sqrt{1-\frac{\Lambda QP}{2\pi}}\right).  \ee
We left the value of the electric field in an implicit form, and took the minus sign in $v_{2}$ in order to write the entropy. When we Taylor expand for small values of the cosmological constant, we recover the previous result for zero potential when $\Lambda\rightarrow 0$ only with the negative sign.
\begin{itemize}
\item $V(\phi)=\beta W(\phi)$.
\end{itemize}
For this case, (\ref{a4}) gives directly
\be W^{2}(u_{D})=\frac{Q^{2}}{(8\pi)^{2}}\frac{1}{(P^{2}+\beta v_{2}^{2})}. \ee
Replacing this in (\ref{a3}) we find $v_{2}$ and all the rest can be found after some algebraic manipulation
\be W(u_{D})=\frac{Q}{8\pi P}\left(1-\frac{\beta Q^{2}}{(8\pi)^{2}}\right)^{1/2} \,\, e=\frac{P}{\left(1-\frac{\beta Q^{2}}{(8\pi)^{2}}\right)^{3/2}} \ee
\be  v_{1}=\frac{QP}{8\pi}\frac{1}{\left(1-\frac{\beta Q^{2}}{(8\pi)^{2}}\right)^{3/2}}, \,\, v_{2}=\frac{QP}{8\pi}\frac{1}{\left(1-\frac{\beta Q^{2}}{(8\pi)^{2}}\right)^{1/2}}, \,\, \label{dirprop}  \ee
\be {\mathcal{E}}=\frac{2\pi QP}{\left(1-\frac{\beta Q^{2}}{(8\pi)^{2}}\right)^{1/2}}. \ee

\begin{itemize}
\item $V(\phi)=\frac{\beta} {W(\phi)}$.
\end{itemize}
This case is a bit simpler to be obtained. The solutions are
\be W(u_{D})=\frac{Q}{8\pi P}\frac{1}{\left(1-\beta P^{2}\right)^{1/2}} \,\, e=\frac{P}{\left(1-\beta P^{2}\right)^{1/2}} \ee
\be  v_{1}=\frac{QP}{8\pi}\frac{1}{\left(1-\beta P^{2}\right)^{3/2}}, \,\, v_{2}=\frac{QP}{8\pi}\frac{1}{\left(1-\beta P^{2}\right)^{1/2}}, \label{invprop}  \ee
\be {\mathcal{E}}=\frac{2\pi QP}{\left(1-\beta P^{2}\right)^{1/2}}. \ee

Notice that we can achieve the zero potential case by setting the constant $\beta$ to zero. Notice also that the $AdS_{2}$ and $S^{2}$ radii, $v_{1}$ and $v_{2}$, are related in equations (\ref{dirprop}) and (\ref{invprop}), by the exchange $Q/(8\pi)\leftrightarrow P$. Of course, one can try other kinds of potentials containing different combinations of these two examples. 

It is good to emphasize again that none of the results depend on the functional form of the coupling $W(\phi)$, and, in order to obtain the black hole entropy, we have assumed that the scalar potential depends on this function in a specific form. These cases illustrate some possible examples that are relevant for gauged supergravities for instance, since the coupling to the field strength is a function of the dilaton field, and the potential may be written as being a function of these couplings. In the next section we show how this works through an explicit computation for the black hole entropy for a specific theory in $AdS_{4}$ space.

\section{$AdS_{4}$ static dyonic black holes}
We apply these ideas to the case of $AdS_{4}$ static black holes. Such system was studied in \cite{Morales:2006gm} by Morales and Samtleben via entropy function, when the gauge field has only electric charge. The solution found by the authors is written in an implicit form, depending on the values of a set of parameters rather than the electric charges. Also, the entropy function was used as a criteria to prove the existence of regular extremal black holes with finite horizon area for a purely electric system \cite{Anabalon:2013sra}. Here, we consider both electric and magnetic charges, meaning that we have a dyonic black hole, and also in the presence of a potential. The notation is explained in the appendix, as the definition of the field $X_{I}$ may change a lot in the literature, and also the definition of the scalar potential. The theory is the $U(1)^{4}$ gauged supergravity in four dimensions, which follows from a truncation of the maximal ${\mathcal{N}=8}, \,\, SO(8)$ supergravity down to the Cartan subgroup of $SO(8)$. The bosonic action is given by \cite{Duff:1999gh}
\be S=\int d^{4}x \sqrt{-g}\left[R-\frac{1}{32}\left(3\sum_{I=1}^{4}(\partial_{\mu} \lambda_{I})^{2}-2\sum_{I<J}\partial_{\mu} \lambda_{I}\partial^{\mu} \lambda_{J}\right)-\frac{1}{4}\sum_{I=1}^{4}X_{I}^{2}(F_{\mu\nu}^{I})^{2} -V(X) \right],\label{actionads4} \ee
where $I=1,...,4$, $16\pi G=1$, and
\be F_{\mu\nu}^{I}=\partial_{\mu}A_{\nu}^{I}-
\partial_{\nu}A_{\mu}^{I}, \,\,\, V(X)=-\frac{g^{2}}{4}\sum_{I<J}\frac{1}{X_{I}X_{J}}, \,\,\, X_{1}X_{2}X_{3}X_{4}=1. \label{definitions}\ee
Notice that we can set the potential term to zero by choosing $g=0$. The near horizon fields will be
\be X_{I}=u_{I}, \,\, F^{I}_{rt}=e^{I}, \,\, F^{I}_{\theta\phi}=p^{I}\sin\theta.  \ee
As the kinetic terms give zero contribution to the entropy function, one can easily do some identifications and add a summation in the entropy function (\ref{neef2}) in order to obtain
\be \mathcal{E}=2\pi \left[ e_{I}q^{I}-4\pi v_{1}v_{2}\left(-\frac{2}{v_{1}}+\frac{2}{v_{2}}+\frac{1}{2}\sum_{I=1}^{4}u_{I}^{2}\left(\frac{e_{I}^{2}}{v_{1}^{2}}-\frac{p_{I}^{2}}{v_{2}^{2}}\right)+4g^{2}\sum_{I<	J}^{4}u_{I}u_{J}\right)\right]. \ee
The attractor equations (\ref{attracteqs}) are 
\be \frac{e_{I}}{v_{1}}= \frac{q^{I}}{(4\pi)u_{I}^{2}v_{2}},  \label{eads}\ee
\be 2-\frac{v_{2}}{2}\sum_{I=1}^{4}u_{I}^{2}\left(\frac{e_{I}^{2}}{v_{1}^{2}}+\frac{p_{I}^{2}}{v_{2}^{2}}\right)+v_{2}V(u)=0,  \label{a1ads}\ee 
\be -2+\frac{v_{1}}{2}\sum_{I=1}^{4}u_{I}^{2}\left(\frac{e_{I}^{2}}{v_{1}^{2}}+\frac{p_{I}^{2}}{v_{2}^{2}}\right)+v_{1}V(u)=0,  \label{a2ads}\ee 
\be u_{I}\left(\frac{e_{I}^{2}}{v_{1}^{2}}-\frac{p_{I}^{2}}{v_{2}^{2}}\right)-\frac{\partial V(u)}{\partial u_{I}}=0. \label{a3ads}\ee
\begin{itemize}
\item V(X)=0.
\end{itemize}
We first discuss the case in which the potential is null. Notice that, if we also set the magnetic charges equal to zero, equation (\ref{a3ads}) will give zero dilaton or zero gauge field on the horizon, and the attractor equations give a trivial solution. This means that for zero potential and zero magnetic charges the attractor equations do not make sense. For nontrivial magnetic charges, one can easily find the solutions:
\be e_{I}=p_{I}, \,\, u_{I}^{2}=\frac{q^{I}}{4\pi p_{I}}, \,\, v_{1}=v_{2}=\frac{1}{2(4\pi)}\sum_{I=1}^{4}q^{I}p_{I}, \label{fieldsADS}\ee
\be \mathcal{E}=2\pi\sum_{I=1}^{4}q^{I}p_{I}. \label{BHADS} \ee
From the results of the previous section we see that this is expected, as we have done nothing but add indices to our fields.

\begin{itemize}
\item $V(X)=-\frac{g^{2}}{4}\sum_{I< J}^{4}\frac{1}{X_{I}X_{J}}$.
\end{itemize}
For this case, equation (\ref{a3ads}) gives
\be \sum_{I=1}^{4}u_{I}^{2}\frac{e_{I}^{2}}{v_{1}^{2}}=\sum_{I=1}^{4}u_{I}^{2}\frac{p_{I}^{2}}{v_{2}^{2}}-2V(u), \label{intads} \ee
and replacing this directly in (\ref{a1ads}) we obtain a quadratic equation
\be v_{2}^{2}V(u)+v_{2}-\frac{1}{2}\sum_{I=1}^{4}u_{I}^{2}p_{I}^{2}=0,\label{eqv2ads} \ee
whose solution is 
\be v_{2}=\frac{-1\pm \sqrt{1+2V(u)\sum_{I=1}^{4}u_{I}^{2}p_{I}^{2}}}{V(u)}. \ee
In order to check if this result has the correct limit we expand it for $V(u)\rightarrow 0$, giving
\be v_{2}=\frac{1}{V(u)}\left(-1\pm\left( 1+V(u)\sum_{I=1}^{4}u_{I}^{2}p_{I}^{2}+{\mathcal{O}(V(u)^{2})}\right)\right).  \ee
Taking the plus sign we can recover the previous case setting the potential to zero. This is again an indication that the case for zero potential should be recovered from this case by setting the potential to zero, i.e. $g^{2}\rightarrow 0$.

Inserting also (\ref{intads}) in (\ref{a2ads}) we find directly
\be v_{1}=\frac{2v_{2}^{2}}{\sum_{I=1}^{4}u_{I}^{2}p_{I}^{2}}. \ee
Also, from (\ref{eads}) we can obtain
\be \sum_{I=1}^{4}u_{I}^{2}\frac{e_{I}^{2}}{v_{1}^{2}}=\sum_{I=1}^{4}\left(\frac{{q^{I}}}{4\pi}\right)^{2}\frac{1}{u_{I}^{2}v_{2}^{2}}. \ee
Inserting this in (\ref{intads}) we have the following equation
\be \sum_{I=1}^{4}\left(\frac{{q^{I}}}{4\pi}\right)^{2}\frac{1}{u_{I}^{2}}=\sum_{I=1}^{4}u_{I}^{2}p_{I}^{2} -2v_{2}^{2}V(u).\label{eqv2ads2}\ee
We can eliminate the last term containing $v_{2}^{2}$ using (\ref{eqv2ads}), and obtain 
\be v_{2}=\frac{1}{2}\sum_{I=1}^{4}\left(\frac{{q^{I}}}{4\pi}\right)^{2}\frac{1}{u_{I}^{2}}. \label{v2dil2}\ee
In order to have a solution for the attractor equations one should use (\ref{v2dil2}) and replace $v_{2}$  in (\ref{eqv2ads}) or in (\ref{eqv2ads2}), and then solve it for $u_{I}$, i.e. write $u_{I}$ in terms of the electric and magnetic charges. It turns out that finding solutions for the resulting equations is a non-trivial task due to the amount of scalar fields and all summations involving them. Based in the analysis made in the previous section, we know that the black hole entropy obtained in (\ref{BHADS}) should be recovered when the potential is set to zero for this case, and so, all other parameters obtained here should  also match the ones obtained before in (\ref{fieldsADS}) in the same limit. By direct observation of the value of the general coupling $W(u_{D})$ on the horizon of the black hole obtained in all the previous cases, we are then led to consider a solution for the dilaton, given by
\be u_{I}^{2}=\frac{q^{I}}{4\pi p_{I}}F(q,p)^{1/2}, \label{ansu}\ee
where $F(q,p)$ is a function of the charges that will be fixed by the attractor equations. With this solution we obtain for the other constant fields 
\be v_{2}=\frac{1}{2(4\pi)}\frac{1}{F(q,p)^{1/2}}\sum_{I=1}^{4}q^{I}p_{I}, \label{ansv2}\ee
\be v_{1}=\frac{1}{2(4\pi)}\frac{1}{F(q,p)^{3/2}}\sum_{I=1}^{4}q^{I}p_{I}, \label{ansv1}\ee
\be e_{I}=\frac{p_{I}}{F(q,p)^{3/2}}. \label{anse}\ee
They will have the correct limit if, at zero potential, the function $F(q,p)\rightarrow 1$. In order to obtain this function, we insert these possible solutions into the attractor equations. It is easy to check that  (\ref{eads}) gives an identity. Equation (\ref{a1ads}), (\ref{a2ads}), (\ref{a3ads}), lead to the same equation. This shows the correctness of our solution (\ref{ansu}). The resulting equation is
\be F(q,p)^{2}-F(q,p)+\frac{g^{2}}{8}\left(\sum_{I=1}^{4}q^{I}p_{I}\right)\left(\sum_{J< K}\sqrt{\frac{p_{J}p_{K}}{q^{J}q^{K}}}\right)=0,  \ee
whose solution is 
\be F(q,p)=\frac{1}{2}\left(1\pm\sqrt{1-\frac{g^{2}}{2}\left(\sum_{I=1}^{4}q^{I}p_{I}\right)\left(\sum_{J< K}\sqrt{\frac{p_{J}p_{K}}{q^{J}q^{K}}}\right)}\right). \label{fg}\ee
In the limit of zero potential we must have $g\rightarrow 0$, and, as alluded before, by consistency $F(q,p)\rightarrow 1$, so the positive sign is the correct one. So, the entropy of the extremal black hole in the presence of electric and magnetic charge for this theory is 
\be \mathcal{E}=2\pi\left(\sum_{I=1}^{4}q^{I}p_{I}\right)
\left[\frac{1}{2}\left(1+\sqrt{1-\frac{g^{2}}{2}\left(\sum_{I=1}^{4}q^{I}p_{I}\right)\left(\sum_{J< K}\sqrt{\frac{p_{J}p_{K}}{q^{J}q^{K}}}\right)}\right)\right]^{-1/2} . \label{entropydepg}\ee
As the dilaton fields are not all independent we can also use  the constraint for the scalar fields,  $X_{1}X_{2}X_{3}X_{4}=1$, and compute the function $F(q,p)$ in a simpler  but not explicit way. This is written as
\be F(q,p)=(4\pi)^{2}\sqrt{\frac{p^{1}p^{2}p^{3}p^{4}}{q^{1}q^{2}q^{3}q^{4}}},\label{fcharges}\ee
and the entropy will be
\be \mathcal{E}=\frac{1}{2}\left(\sum_{I=1}^{4}q^{I}p_{I}\right)\left(\frac{q^{1}q^{2}q^{3}q^{4}}{p^{1}p^{2}p^{3}p^{4}}\right)^{1/4} .  \ee
Written in this form the entropy does not show explicitly the coupling constant $g^{2}$. One can obtain it as a function of the charges of the black hole by using (\ref{fg}) and (\ref{fcharges}), i.e. 
\be g^{2}=\frac{8(4\pi)^{4}\left(\sqrt{\frac{p^{1}p^{2}p^{3}p^{4}}{q^{1}q^{2}q^{3}q^{4}}}-\frac{p^{1}p^{2}p^{3}p^{4}}{q^{1}q^{2}q^{3}q^{4}}\right)}{\left(\sum_{I=1}^{4}q^{I}p_{I}\right)\left(\sum_{J< K}\sqrt{\frac{p_{J}p_{K}}{q^{J}q^{K}}}\right)}. \label{coupling} \ee
This means that, for this geometry, the charges must be chosen in such a way to satisfy this specific condition  on the horizon  of the extremal black hole. This is due to the fact that, for gauged supergravities, in order to obtain a consistent truncation, the scalar fields should be constrained (like $X_{1}X_{2}X_{3}X_{4}=1$ in the $U(1)^{4}$ case), and this constraints the values of the electric and magnetic charges on the horizon of the extremal black hole, since $g^{2}$ is a parameter defining the theory that one can freely vary. In other words, (\ref{coupling}) is a condition on the charges of the black hole, and not on the coupling constant $g^{2}$.  In the absence of such constraint the electric and magnetic charges can also be chosen freely.

If the dyonic black hole solution is invariant under duality transformation, $u_{I}\rightarrow u_{I}^{-1}$ and $(q_{I}/4\pi) \leftrightarrow p^{I}$, then the dilaton field is invariant, since $F\rightarrow F^{-1}$\footnote{As was pointed out in \cite{Duff:1999gh}, for the case of purely electric or purely magnetic black holes, there is a direct correspondence between the supersymmetry properties of the electric and magnetic solutions in the absence of gauging ($g=0$), which apparently does not happen for the gauged theory. However it is not clear yet if the dyonic solution of this theory preserves the supersymmetry properties under electric-magnetic duality.  Here we just take the dual in order to show that it is possible to recover the entropy computed for purely magnetic black holes.}. But notice that the entropy changes to 
\be \mathcal{E}=8\pi^{2}\left(\sum_{I=1}^{4}q^{I}p_{I}\right)\left(\frac{p^{1}p^{2}p^{3}p^{4}}{q^{1}q^{2}q^{3}q^{4}}\right)^{1/4} .  \ee
If one makes the rescaling $q^{I}\equiv \epsilon$, then the electric charge disappears from this formula, and the result is just 
\be \mathcal{E}=8\pi^{2}\left(\sum_{I=1}^{4}p_{I}\right)\left(p^{1}p^{2}p^{3}p^{4}\right)^{1/4}. \label{entropymag}\ee
In order to obtain the same result from equation (\ref{entropydepg}), one should take the limit of  magnetic black holes, i.e. $\epsilon \rightarrow 0$. A formula for purely magnetic black hole was found for an $\mathcal{N}=2$ supersymmetric theory  in reference \cite{Cacciatori:2009iz}, which is given by
\be S=2\pi^{2}\left(\prod_{I=1}^{4}\frac{p^{I}}{g_{I}} \right)^{1/4},\label{entropymag1}\ee
where $g_{I}=g\xi_{I}$, for some constant $\xi_{I}$. As pointed out in the same reference, such a model can be embedded in $\mathcal{N}=8$ gauged supergravity. The gravitino is charged in gauged supergravity, and so, for topological consistency in a magnetic backgroung field, the charges should satisfy the Dirac quantization condition: for BPS configurations this is just $\sum g_{I}p^{I}\in \kappa$, where $\kappa$ is the curvature of the horizon geometry. In the case of an $S^{2}$ horizon we have $\kappa=1$. For the $U(1)^{4}$ theory the Dirac quantization is written as $-8\pi \sum g_{I}p^{I}=\mathbb{Z}$. Using this condition, it is easy to see that equations (\ref{entropymag}) and (\ref{entropymag1}) agree for some choice of constant $\xi_{I}$, which shows that one can recover the purely electric case from the dyonic one, if the dyonic solution is invariant under electric-magnetic duality in the presence of gaugings. It should be emphasized that  the Sen's formalism allows us to obtain information about the fields in the near horizon for general extremal black holes: they do not necessarily need to be BPS, and this is the reason why we should use $-8\pi \sum g_{I}p^{I}=\mathbb{Z}$, instead of the Dirac quantization for BPS black holes.

The question that naturally arises in this work is whether solutions of the kind (\ref{ansu}) represent a general feature of the scalar fields for any theory of the kind (\ref{ad}). All the results obtained here depended strongly on the functional form of the dilaton potential. Some complications in the equations might arise when the potential is a complicated function (like logarithmic or exponential)of the couplings to the field strength ($W(\phi)$ in (\ref{ad}) and $u_{I}^{2}$ in (\ref{actionads4})). Potentials having polynomial functional forms of these couplings are the most common ones in gauged supergravities, and, at least in $4$ spacetime dimensions, it seems that (\ref{ansu}) will allow one to solve the attractor equations for any of these theories. Of course, one should check it case by case, since the results presented here are only an illustration of why this should work, but not a complete proof. Also, for more general theories in $D$-dimensions containing Chern-Simons terms and higher order derivatives, like $RF^{2}$ and $R^{2}$ for instance, the attractor equations might present higher powers of $v_{1}$ and $v_{2}$, and such solutions are much more difficult to be obtained, but of course, in the limit when the coupling constants of such terms go to zero, we must recover the case for zero potential. How our results change in such cases may be a subject of future work.

\section{Summary and conclusions}
We have shown that Sen's entropy function allows us to obtain the entropy for the extremal dyonic black hole with near horizon geometry given by $AdS_{2}\times S^{2}$, whenever the action is of the type (\ref{ad}) with the potential having a specific (polynomial) form. After analysing some general examples, we have shown explicitly how to solve the attractors the equations for the case of $U(1)^{4}$ gauged supergravity in $4$ dimensions. One should write all the attractor equations in terms of the dilaton fields, and then assume that the dilaton fields on the horizon are given by
\be u_{I}^{2}=\frac{q^{I}}{4\pi p_{I}}F(q,p)^{1/2}.\label{final}\ee
Then, after inserting this solution in the attractor equations, one solves the attractor equations for $F(q,p)$ and obtain the solution for all the fields, $u_{I}$, $v_{1}$, $v_{2}$ and $e_{I}$. The function $F(q,p)$ may not be unique in the sense that it is a solution of a polynomial equation (quadratic equation for $U(1)^{4}$ case). The solution that must be chosen is the one that satisfies  
\be \lim_{V\rightarrow 0} F(q,p)= 1.  \ee
One can see that this is achieved in (\ref{fg}) by choosing the $+$ sign.  This shows that the scalar attractors change in a specific way that depends crucially on the functional form of the potential, and we speculated about (\ref{final}) being a universal feature of the sort of theories treated here. We emphasize that this should be understood as a parametrization,  since we have just singled out a factor (or factors) of $\frac{q}{4\pi p}$ from the function $F(q,p)$, and not an ansatz. No restriction on the function $F(q,p)$ was made. As the attractor equations are a system with the same number of variables and linearly independent equations, the system is completely determined.  

We have obtained the values of the dilaton and the electric field on the horizon of the extremal dyonic black hole for the $U(1)^{4}$ gauged supergravity theory in $4$ dimensions. We have also derived explicitly the near horizon metric and the entropy  of these black holes, and showed that the constraint $X_{1}X_{2}X_{3}X_{4}=1$ fixes the electric and magnetic charges on the horizon to satisfy (\ref{coupling}) (this may be associated to the mass $m^{2}$ of the dilaton field for some potentials). Through the $U(1)^{4}$ example, we also showed the power of the entropy function formalism in determining the near horizon metric and the entropy for dyonic black holes. This approach might give some indications about how the full metric for the non-extremal dyonic dilatonic black hole should be, since the near horizon metric computed here should be obtained by taking the extremal limit in the full non-extremal solution. None of these results could have been obtained without assuming that the scalars on the horizon are written as (\ref{final}).

{\bf{Acknowledgments}}\\
The author would like to thank Thiago R. Araujo, Johanna Erdmenger, Daniel Fernandez and Horatiu Nastase for reading the final version of this paper and for useful discussions. The author is grateful for the hospitality of the Max-Planck-Institut f\"{u}r  Physik (Werner Heisenberg Institut), where part of this work was developed. The work of PG is supported by FAPESP grant 2013/00140-7 and 2015/17441-5.

\appendix

\section{$U(1)^{4}$ gauged supergravity in 4 dimensions}
Due to some different notation in the definition of the dilaton field, we present a brief review to explain our definitions for the case of $U(1)^{4}$ supergravity in 4 dimensions. The bosonic action with the same field definition and coefficients given in \cite{Duff:1999gh} is
\bea S=\int d^{4}x \sqrt{-g}\left[R-\frac{1}{2}\left((\partial \phi^{(12)})^{2}+(\partial \phi^{(13)})^{2}+(\partial \phi^{(14)})^{2}\right)-V\right. \nonumber \\
\left. -2\left(e^{-\lambda_{1}}(F_{\mu\nu}^{(1)})^{2}+e^{-\lambda_{2}}(F_{\mu\nu}^{(2)})^{2}+e^{-\lambda_{3}}(F_{\mu\nu}^{(3)})^{2}+e^{-\lambda_{4}}(F_{\mu\nu}^{(4)})^{2}\right)  \right], \eea
where the scalar combinations ${\lambda}$ are given by
\begin{equation} \begin{array}{ccc}
\lambda_{1} & = & -\phi^{(12)}-\phi^{(13)}-\phi^{(14)},\\
\lambda_{2} & = & -\phi^{(12)}+\phi^{(13)}+\phi^{(14)},\\
\lambda_{3} & = & \;\;\, \phi^{(12)}-\phi^{(13)}+\phi^{(14)},\\
\lambda_{4} & = & \;\;\,\phi^{(12)}+\phi^{(13)}-\phi^{(14)},\\
\end{array} \end{equation}
and the scalar potential is 
\be V=-4g^{2}\left(\cosh\phi^{(12)}+\cosh\phi^{(13)}+
\cosh\phi^{(14)}\right). \ee
Notice that the scalars are not all independent,
\be \lambda_{1}+\lambda_{2}+\lambda_{3}+\lambda_{4}=0. \ee
We can write the scalar $\phi^{(ij)}$ in terms of the fields $\lambda$ as
\be \phi^{(12)}=\frac{1}{4}(-\lambda_{1}-\lambda_{2}+\lambda_{3}+
\lambda_{4}), \ee
\be \phi^{(13)}=\frac{1}{4}(-\lambda_{1}+\lambda_{2}-\lambda_{3}+
\lambda_{4}), \ee
\be \phi^{(14)}=\frac{1}{4}(-\lambda_{1}+\lambda_{2}+\lambda_{3}-
\lambda_{4}). \ee
We can rewrite parts of the kinetic terms as
\begin{eqnarray} \partial_{\mu}\phi^{12}\partial^{\mu}\phi^{12}=\frac{1}{16}\left[(\partial_{\mu}\lambda_{1})^{2}+
(\partial_{\mu}\lambda_{2})^{2}+(\partial_{\mu}
\lambda_{3})^{2} +(\partial_{\mu}\lambda_{4})^{2}\right.\nonumber \\
+2\partial_{\mu}\lambda_{1}\partial^{\mu}\lambda_{2}
-2\partial_{\mu}\lambda_{1}\partial^{\mu}\lambda_{3}
-2\partial_{\mu}\lambda_{1}\partial^{\mu}\lambda_{4}
\nonumber \\ \left.
-2\partial_{\mu}\lambda_{2}\partial^{\mu}\lambda_{3}
-2\partial_{\mu}\lambda_{2}\partial^{\mu}\lambda_{4}
+2\partial_{\mu}\lambda_{3}\partial^{\mu}\lambda_{4}
\right],\end{eqnarray}
\begin{eqnarray} \partial_{\mu}\phi^{13}\partial^{\mu}\phi^{13}=\frac{1}{16}\left[(\partial_{\mu}\lambda_{1})^{2}+
(\partial_{\mu}\lambda_{2})^{2}+(\partial_{\mu}
\lambda_{3})^{2} +(\partial_{\mu}\lambda_{4})^{2}\right.\nonumber \\
-2\partial_{\mu}\lambda_{1}\partial^{\mu}\lambda_{2}
+2\partial_{\mu}\lambda_{1}\partial^{\mu}\lambda_{3}
-2\partial_{\mu}\lambda_{1}\partial^{\mu}\lambda_{4}
\nonumber \\ \left.
-2\partial_{\mu}\lambda_{2}\partial^{\mu}\lambda_{3}
+2\partial_{\mu}\lambda_{2}\partial^{\mu}\lambda_{4}
-2\partial_{\mu}\lambda_{3}\partial^{\mu}\lambda_{4}
\right],\end{eqnarray}
\begin{eqnarray} \partial_{\mu}\phi^{14}\partial^{\mu}\phi^{14}=\frac{1}{16}\left[(\partial_{\mu}\lambda_{1})^{2}+
(\partial_{\mu}\lambda_{2})^{2}+(\partial_{\mu}
\lambda_{3})^{2} +(\partial_{\mu}\lambda_{4})^{2}\right.\nonumber \\
-2\partial_{\mu}\lambda_{1}\partial^{\mu}\lambda_{2}
-2\partial_{\mu}\lambda_{1}\partial^{\mu}\lambda_{3}
+2\partial_{\mu}\lambda_{1}\partial^{\mu}\lambda_{4}
\nonumber \\ \left.
+2\partial_{\mu}\lambda_{2}\partial^{\mu}\lambda_{3}
-2\partial_{\mu}\lambda_{2}\partial^{\mu}\lambda_{4}
-2\partial_{\mu}\lambda_{3}\partial^{\mu}\lambda_{4}
\right].\end{eqnarray}
The full kinetic term is
\begin{eqnarray}
 (\partial_{\mu}\phi^{12})^{2}+
(\partial_{\mu}\phi^{13})^{2}+
(\partial_{\mu}\phi^{14})^{2}=\frac{1}{16}\left[3(\partial_{\mu}\lambda_{1})^{2}+
3(\partial_{\mu}\lambda_{2})^{2}+3(\partial_{\mu}
\lambda_{3})^{2} \right.\nonumber \\
+3(\partial_{\mu}\lambda_{4})^{2} -2\partial_{\mu}\lambda_{1}\partial^{\mu}\lambda_{2}
-2\partial_{\mu}\lambda_{1}\partial^{\mu}\lambda_{3}
-2\partial_{\mu}\lambda_{1}\partial^{\mu}\lambda_{4}
-2\partial_{\mu}\lambda_{2}\partial^{\mu}\lambda_{3}
\nonumber \\ \left.
-2\partial_{\mu}\lambda_{2}\partial^{\mu}\lambda_{4}
-2\partial_{\mu}\lambda_{3}\partial^{\mu}\lambda_{4}
\right]. \end{eqnarray}
The potential term is written as
\be V=-2g^{2}\left(e^{\phi^{12}}+e^{\phi^{13}}+
e^{\phi^{14}}+\frac{1}{e^{\phi^{12}}}+\frac{1}{e^{\phi^{13}}}+\frac{1}{e^{\phi^{14}}}\right), \ee
and in terms of the fields $\lambda$ we have
\be V=-2g^{2}\left[e^{\frac{1}{2}(\lambda_{1}+\lambda_{2})}+e^{\frac{1}{2}(\lambda_{1}+\lambda_{3})}+e^{\frac{1}{2}(\lambda_{1}+\lambda_{4})}+e^{\frac{1}{2}(\lambda_{2}+\lambda_{3})}+e^{\frac{1}{2}(\lambda_{2}+\lambda_{4})}+e^{\frac{1}{2}(\lambda_{3}+\lambda_{4})}\right]. \ee
In order to have the Maxwell term written with a factor $1/4$, we  redefine the exponential of the fields as
\be \frac{X_{I}}{\sqrt{8}}\equiv e^{-\frac{\lambda_{I}}{2}}\Rightarrow \partial_{\mu}\lambda_{I}=-2\frac{\partial_{\mu}X_{I}}{X_{I}}. \ee
The potential is then
\be V=-\frac{g^{2}}{4}\sum_{I< J}\frac{1}{X_{I}X_{J}}.  \ee
The action is finally rewritten as
\be S=\int d^{4}x \sqrt{-g}\left[R-\frac{1}{32}\left(3\sum_{I=1}^{4}(\partial_{\mu} \lambda_{I})^{2}-2\sum_{I<J}\partial_{\mu} \lambda_{I}\partial^{\mu} \lambda_{J}\right)-\frac{1}{4}\sum_{I=1}^{4}X_{I}^{2}(F_{\mu\nu}^{I})^{2} -V \right]. \ee

\providecommand{\href}[2]{#2}\begingroup\raggedright\endgroup

\end{document}